\documentclass[preprint,12pt]{elsarticle}

\usepackage{amssymb}
\usepackage{algorithm}
\usepackage{algpascal}
\usepackage{algpseudocode}
\usepackage{etoolbox}

 \biboptions{sort&compress}

\algtext*{EndWhile}% Remove "end while" text
\algtext*{EndIf}% Remove "end if" text

\begin{document}

\pagestyle{empty}
\begin{frontmatter}

%% Title, authors and addresses

%% use the tnoteref command within \title for footnotes;
%% use the tnotetext command for theassociated footnote;
%% use the fnref command within \author or \address for footnotes;
%% use the fntext command for theassociated footnote;
%% use the corref command within \author for corresponding author footnotes;
%% use the cortext command for theassociated footnote;
%% use the ead command for the email address,
%% and the form \ead[url] for the home page:
%% \title{Title\tnoteref{label1}}
%% \tnotetext[label1]{}
%% \author{Name\corref{cor1}\fnref{label2}}
%% \ead{email address}
%% \ead[url]{home page}
%% \fntext[label2]{}
%% \cortext[cor1]{}
%% \address{Address\fnref{label3}}
%% \fntext[label3]{}

\title{A new algorithm for Many to Many Matching with Demands and Capacities}

%% use optional labels to link authors explicitly to addresses:
%% \author[label1,label2]{}
%% \address[label1]{}
%% \address[label2]{}

\author[la1]{Fatemeh Rajabi-Alni}

\corref{cor1}\ead {fatemehrajabialni@yahoo.com}
\cortext[cor1]{Corresponding author.}

 \address[la1]{Department of Computer Engineering, Islamic Azad University,\\North Tehran Branch, Tehran, Iran.\fnref{label3}}

\begin{abstract}
%% Text of abstract
Let $A=\{a_1,a_2,\dots,a_s\}$ and $\{b_1,b_2,\dots,b_t\}$ with $s+r=n$, \textit {the many to many point matching with demands and capacities} matches each point $a_i \in A$ to at least $\alpha_i$ and at most ${\alpha '}_i$ points in $B$, and each point $b_j \in B$ to at least $\beta_j$ and at most ${\beta '}_j$ points in $A$ for all $1 \leq i \leq s$ and $1 \leq j \leq t$. In this paper, we present an $O(n^4)$ time and $O(n)$ space algorithm for this problem.
\end{abstract}

\begin{keyword}
%% keywords here, in the form: keyword \sep keyword

%% MSC codes here, in the form: \MSC code \sep code
%% or \MSC[2008] code \sep code (2000 is the default)
many to many matching\sep Hungarian method\sep bipartite graph\sep points with demands and capacities

\end{keyword}

\end{frontmatter}

%%
%% Start line numbering here if you want
%%
% \linenumbers

\section{Introduction}

A \textit {matching} between two sets defines a relationship between their elements. The matching is used in various fields such as computational biology \cite{1}, pattern recognition \cite{2}, computer vision \cite{3}, music information retrieval \cite{4}, and computational music theory \cite{5}. 
A \textit {many-to-many matching} between $A$ and $B$ assigns each point in $A$ to one or more points in $B$, and vise versa.

Let $A$ and $B$ be two sets with $|A|+|B|=n$, Eiter and Mannila \cite{6} proposed an $O(n^3)$ algorithm for the minimum many-to-many matching problem between $A$ and $B$ by reducing the problem to the minimum-weight perfect matching problem in a bipartite graph. 

\textit {The minimum many-to-many matching with demands and capacities}, here called \textit {MMDC} matching, is a matching in which each point $a_i\in A$ is matched to at least $\alpha_i$ and at most ${\alpha '}_i$ points in $B$, and each point $b_j\in B$ is matched to at least $\beta_j$ and at most ${\beta '}_j$ points in $A$, such that sum of the matching costs is minimized.
Schrijver \cite{7} solved the MMDC matching problem in strongly polynomial time. In this paper, we present a new algorithm that computes an MMDC matching between $A$ and $B$ in $O(n^4)$ time using $O(n)$ space. In section \ref{Preliminaries}, we review the basic Hungarian algorithm and some preliminary definitions. In section \ref{newalgorithms}, we present our new algorithm. 

\section {Preliminaries}
\label{Preliminaries}
Given an undirected bipartite graph $G=(A \cup B, E)$, A \textit {maximum matching} $M$ is a matching that for any other matching $M'$, we have $Weight(M') < Weight(M)$. A path with the edges alternating between $M$ and $E-M$ is called an \textit {alternating path}. Each vertex $v$ that is incident to one edge in $M$ is called a \textit {matched vertex}; otherwise it is a \textit {free vertex}. An alternating path that its both endpoints are free is called an \textit {augmenting path}. Note that if the $M$ edges of an augmenting path is replaced with the $E-M$ ones, its size increases by $1$. Let $V=A \cup B$, a \textit {vertex labeling function} $l: V \rightarrow R$ assigns a label to each vertex $v \in V$. A vertex labeling that in which $l(a)+l(b) \ge Weight(a,b)$ for all $a \in A$ and $b \in B$ is called a \textit {feasible labeling}. The equality graph of a feasible labeling $l$ is a graph $G=(V,E_l)$ such that $E_l=\{(a,b)| l(a)+l(b)=Weight(a,b)\}$. The \textit {neighbors} of a vertex $u \in V$ is defined as $N_l(u)=\{v| (v,u) \in E_l\}$. Consider a set of the vertices $S \subset V$, the neighbors of $S$ is $N_l(S)=\bigcup_{u \in S} N_l(u)$.
\newtheorem{lemma}{Lemma}
\begin{lemma}
\label{lem1}
Consider a feasible labeling $l$ of an undirected bipartite graph $G=(A\cup B, E)$ and $S \subset A$  with $T=N_l(S)\neq B$, let 
$$\alpha_l=\min_{a_i \in S, b_j \notin T}\{l(a_i)+l(b_j)-Weight(a_i,b_j)\}.$$ If the labels of the vertices of $G$ is updated such that:
$$l'(v)=\left\{ 
\begin{array}{lr}
l(v)-\alpha_l & if \  v \in S 
 \\ 
 l(v)+\alpha_l & if\  v  \in T 
 \\ 
 l(v) & Otherwise 
 \end{array}
\right.$$ 
then, $l'$ is also a feasible labeling.
\end{lemma}

\textbf {Proof.} Note that $l$ is a feasible labeling, so we have $l(a)+l(b)\ge Weight(a,b)$ for each edge $(a,b)$ of $E$. After the update four cases arise:
\begin{itemize}
\item $a \in S$ and $b \in T$. In this case $$l'(a)+l'(b)=l(a)-\alpha_l+l(b)+\alpha_l=l(a)+l(b)\ge Weight(a,b).$$ 
\item $a \notin S$ and $b \notin T$. We have $$l'(a)+l'(b)=l(a)+l(b)\ge Weight(a,b).$$
\item $a \notin S$ and $b \in T$. We see that $$l'(a)+l'(b)=l(a)+l(b)+\alpha_l>l(a)+l(b)\ge Weight(a,b).$$
\item $a \in S$ and $b \notin T$. In this situation we have 
$$l'(a)+l'(b)=l(a)-\alpha_l+l(b).$$
Two cases arises: 
\begin{itemize}

\item $l(a)+l(b)-Weight(a,b)=\alpha_l$. So 
$$l'(a)+l'(b)=l(a)-\alpha_l+l(b)=l(a)-l(a)-l(b)+Weight(a,b)+l(b)=Weight(a,b).$$ Hence, $E_l \subset E_{l'}$.  

\item $l(a)+l(b)-Weight(a,b)>\alpha_l$. Obviously 
$$l'(a)+l'(b)=l(a)-\alpha_l+l(b)>Weight(a,b).$$

\end{itemize}

\end{itemize}
\qed

\newtheorem{theorem}{Theorem}
\begin{theorem}
If $l$ is feasible labeling and $M$ is a Perfect matching in $E_l$, then $M$ is a max-weight matching \cite{8}.
\end{theorem}
\textbf {Proof.} Suppose that $M'$ is a perfect matching in $G$, since each  vertex is incident to exactly one edge of $M'$ we have:
$$Weight(M')=\sum_{(a,b) \in M'} Weight(a,b)\le \sum_{v \in (A \cup B)}l(v).$$ So $\sum_{v \in (A \cup B)}l(v)$ is an upper bound for each perfect matching. Now assume that $M$ is a perfect matching in $E_l$:
$$Weight(M)=\sum_{e \in M}l(e)=\sum_{v \in (A \cup B)}l(v).$$ It is obvious that $M$ is an optimal matching. 
\qed

In the following, we briefly describe the basic Hungarian algorithm which computes the maximum many to many matching between two sets. The input bipartite graph $G=(A \cup B, E)$ is a complete bipartite graph that in which $|A|=|B|=n$.

\makeatletter
\expandafter\patchcmd\csname\string\algorithmic\endcsname{\itemsep\z@}{\itemsep=0.5ex plus0.5pt}{}{}
\makeatother

\algsetblock[Name]{Initial}{}{3}{1cm}
\alglanguage{pseudocode}
\begin{algorithm}
\caption{The Basic Hungarian algorithm($A$,$B$)}
\begin{algorithmic}[1]
\Initial \Comment Find an initial feasible labeling $l$ and a matching $M$ in $E_l$
\State Let $l(b_j)=0, \ for \ all \ 1 \le j \le t$
\State $l(a_i)=\max_{j=1}^t Weight(a_i,b_j)\  for\  all \ 1 \le i \le s$
\State $M= \emptyset$
\While {$M$ is not perfect}

 \State Select a free vertex $a_i  \in A$ and set $S = \{a_i\}$, $T=\emptyset$
  \Repeat
     \While {$N_l(S)=T$}
        \State Update the labels according to Lemma \ref{lem1}
        \EndWhile
         \State Select $b_j  \in N_l (S)-T$ 
          \If {$b_j$ is not free}\Comment ($b_j$ is matched to the vertex $z$, extend the alternating tree)
            \State $S = S \cup {z},T = T \cup {b_j}$.
           \EndIf
          \Until {$b_j$ is free}   
    \State Augment $M$
\EndWhile
\Return $M$
\end{algorithmic}
 \end{algorithm}

In line $1$, we label all points of $B$ with zero and each point $a_i \in A$ with $\max_{j=1}^n Weight(a_i,b_j)$ to get an initial feasible labeling. Note that $M$ can be empty. It is obvious that for computing the minimum cost many to many matching using the Hungarian algorithm we must weight the edge $(a_i,b_j)$ by $1/Weight(a_i,b_j)$.

\begin{lemma}
Each augmenting path is a 4-vertex path.
\end{lemma}
\textbf {Proof.} Suppose that the lemma is false. Let $p=a_1,b_1,a_2,b_2, \dots, b_k$ be an augmenting path with more than four vertices, that is $k>2$. Note that $a_1$ and $b_k$ are free nodes. It is obvious that the first edge is in $E-M$, so the second , third, and fourth edges of $p$ are in $M$, $E-M$, and $M$, respectively. Since the third edge $(a_2,b_2)$ is in $E-M$, the fourth edge $(b_2,a_3)$ must be in $M$. Note that $b_2$ is a free node. A contradiction. 
\qed

\section{The algorithm}
\label{newalgorithms}
In this section, we describe our new algorithm which is based on the well known Hungarian algorithm. Consider two point sets $A=\{a_1,a_2,\dots,a_s\}$ and $B=\{b_1,b_2,\dots,b_t\}$ with $s+t=n$. Let $D_A=\{\alpha_1,\alpha_2,\dots,\alpha_s\}$ and $D_B=\{\beta_1,\beta_2,\dots,\beta_t\}$ denote the demand sets of $A$ and $B$, respectively. Let $C_A=\{{\alpha '}_1,{\alpha '}_2,\dots,{\alpha '}_s\}$ and $C_B=\{{\beta '}_1,{\beta '}_2,\dots,{\beta '}_t\}$ be the capacity sets of $A$ and $B$, respectively. Without loss of generality, we assume that $\sum_{i=1}^s{\alpha}'_i>\sum_{j=1}^t{\beta}'_j$. 

\begin{theorem}
Let $A$ and $B$ be two sets with $|A|+|B|=n$, an MMDC matching between $A$ and $B$ can be computed in $O(n^4)$ time.
\end{theorem}

\textbf {Proof.} 

\begin{figure}
\vspace{-8cm}
\hspace{-11cm}
\resizebox{2.5\textwidth}{!}{%
  \includegraphics{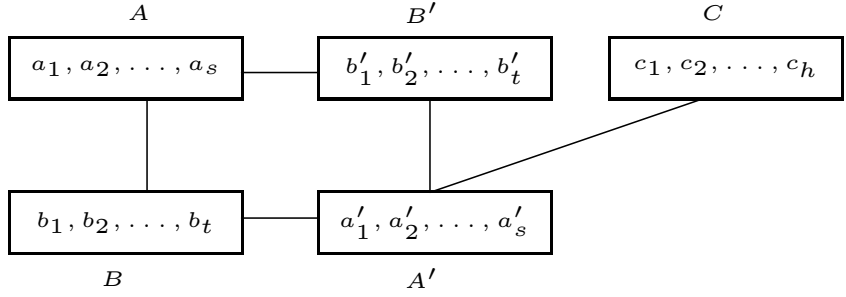}
}
% If not, use
%\vspace{5cm}       % Give the correct figure height in cm
\vspace{-37.5cm}
\caption{Our constructed complete bipartite graph with $h=\sum_{i=1}^s{\alpha}'_i-\sum_{j=1}^t{\beta}'_j$.}
\label{fig:1}       % Give a unique label
\end{figure}

We first construct a bipartite graph as follows. Consider the complete bipartite graph $G=(X \cup Y, E)$ where $X=A \cup A'$ and $Y=B \cup B' \cup C$ (see Figure \ref{fig:1}). 
A \textit {complete connection} between two sets is a connection that in which each element of one set is connected to all elements of the other set. We show each set of the vertices by a rectangle and the complete connection between them by a line connecting the two corresponding rectangles. 

Given $A= \{a_1, a_2, \dots, a_s\}$ and $B= \{b_1, b_2, \dots, b_t\}$, there exists a complete connection between $A$ and $B$ such that the weight of $(a_i, b_j)$ is equal to the cost of matching the point $a_i$ to $b_j$ for all $1 \le i \le s$ and $1 \le j \le t$. Let $B'= \{b'_1, b'_2, \dots, b'_t\}$ and $A'= \{a'_1, a'_2, \dots, a'_s\}$, each point of $A$ is connected to the all points of $B'$ such that the weight of $(a_i,b'j)$ is equal to the weight of $(a_i,bj)$. There exists also a complete connection between the sets $B$ and $A'$ such that the weight of $(a'_i,bj)$ is equal to the weight of $(a_i,bj)$. We have a set $C=\{c_1,c_2,\dots,c_h\}$ that in which $h=\sum_{i=1}^s{\alpha}'_i-\sum_{j=1}^t{\beta}'_j$. In fact, we use $C$ to get $|X|=|Y|$. Each vertex of $A'$ is connected to all vertices of $C$ with zero weighted edges.

Now we apply our new algorithm, Algorithm \ref{MMDC}, on above bipartite graph $G$. Let $Cap(u)$ and $Dem(u)$ denote the capacity and the demand of the vertex $u$; so for all $i,j$ we have $Dem(a_i)=\alpha_i$, $Dem(b_j)=\beta_j$, $Cap(a_i)={\alpha }'_i$, and $Cap(b_j)={\beta }'_j$. 

In our algorithm, a vertex $x$ is free to another vertex $y$ if $x$ is not matched with $y$ in $M$ and has at least one empty capacity.  
So $a_i \in A$ and $a'_i \in A'$ are called free vertices to a vertex $b$ that are not matched with it in $M$, if \newline $Num(a_i)<Dem(a_i)$ and $Num(a'_i)<Cap(a_i)-Dem(a_i)$, respectively. \newline Also the vertices $b_j$ and $b'_j$ are free to another vertex that is not incident in $M$ to them, when \newline $Num(b_j)<Dem(b_j)$ and $Num(b'_j)<Cap(b_j)-Dem(b_j)$, respectively.

In fact, we save the current number of the vertices that are matched to the vertices of $A$, $B$, $A'$, and $B'$ in the arrays $A[1 \dots s]$, $B[1 \dots t]$, $A'[1 \dots s]$, and $B'[1 \dots t]$, respectively; for example $A[i]$ shows the number of the nodes that are matched to $a_i$. The initial values of the arrays is $0$; when a new point is matched to their representing node their values are increased by $1$. Assume that $Num(u)$ returns the number of the vertices that are matched to $u$ so far. So $Num(a_i)=A[i]$, $Num(a'_i)=A'[i]$, $Num(b_j)=B[j]$, and finally $Num(b'_j)=B'[j]$. Note that the procedures $IsFree(u)$ and $IsMatched(u)$ return $True$ if $Num(u)<Cap(u)$ and $Num(u)=Cap(u)$, respectively. So in the augmenting path $a,b,c,d$, $a$ is free to $b$, $b$ is matched to $c$, and $d$ is free to $c$. Now we change the basic Hungarian algorithm as follows.

\algsetblock[Name]{Initialize}{}{4}{1cm}

\alglanguage{pseudocode}
\begin{algorithm}
\caption{The MMDC Hungarian algorithm($DA$, $CA$, $DB$, $CB$)}
\label{MMDC}
\begin{algorithmic}[1]

\Initialize \Comment Find an initial feasible labeling $l$ and a matching $M$ in $E_l$
\State Let $l(b_j), l(b'_j)=0, \ for \ all \ 1 \le j \le t$
\State $l(a_i)=\max_{j=1}^t (\max(Weight(a_i,b_j),Weight(a_i,b'_j))\  for\  all \ 1 \le i \le s$
\State $l(a'_i)=\max_{j=1}^t Weight(a'_i,b_j)$$\  for\  all \ 1 \le i \le s$
\State Let $M=\emptyset$ 
\item[]
   \While{$\{u \in A\cup A',\ with\ IsFree(u) \}\neq \emptyset$}

 \State Select $u  \in A\cup A'$ with $IsFree(u)$ 
 \State Set $S = \{u\}, T =\emptyset$
\item[]
  \Repeat
     \While {$N_l(S)=T$}
        \Statex \Comment Update the labels according to Lemma \ref{lem1}

\State Let $\alpha_l=\min_{s_i \in S, t_j \notin T}\{l(s_i)+l(t_j)-Weight(s_i,t_j)\}$
\State Let
$l'(v)=\left\{ 
\begin{array}{lr}
l(v)-\alpha_l & if \  v \in S
 \\ 
 l(v)+\alpha_l & if\  v  \in T 
 \\ 
 l(v) & Otherwise 
 \end{array}
\right.$ 

        \EndWhile
\item[]
         \State Select $y  \in N_l (S)-T$
          \If {$IsMatched(y)$ }\Comment ($Num(y)=Cap(y)$)
\Statex\Comment ($y$ is matched to some vertices $z$)

            \State $S = S \cup \{z|(z,y) \in M\},T = T \cup \{y\}$.
           \EndIf
          \Until {$IsFree(y)$}   
\item[]
    \State $Augment(M)$
    \EndWhile

\end{algorithmic}
\end{algorithm}

We first label the vertices of our bipartite graph $G$ using an initial feasible labeling in lines $2-4$. Algorithm \ref{MMDC} has a $while$ loop where $O(n^2)$ times iterates and $\sum_{i=1}^s\alpha_i+\sum_{j=1}^t\beta_j$ edges are selected. In each iteration of our algorithm $|M|$ increases by $1$. Let $$slack_y=\min_{x \in S}\{l(x)+l(y)-Weight(x,y)\}.$$ In line $17$ ofAlgorithm \ref{MMDC}, the values of all slacks must be updated when a vertex is moved form $\bar S$ to $S$. This is done in $O(n)$ time. During our algorithm $s+t=n$ vertices are moved from $\bar S$ to $S$, so it takes the total time of $O(n^2)$. 

In lines $11$, we can compute the value of $\alpha_l$ by:
$$\alpha_l=\min_{y \notin T}slack_y,$$
in $O(n)$ time. After computing the value of $\alpha_l$ and updating the labels of the vertices, we must also update the values of the slacks. This can be done using:
$$\forall y \notin T slack_y=slack_y-\alpha_l.$$ 
In each iteration the value of $\alpha_l$ may be computed at most $O(n)$ times, that takes $O(n)$ time each time, so running each iteration takes at most $O(n^2)$ time.
Our algorithm has $O(n^2)$ iteration with $O(n^2)$ time, so it runs in $O(n^4)$ time. 

\qed

\section{Conclusion}
\label{ConclusionSect}
In this paper, we presented an $O(n^4)$ time and $O(n)$ space algorithm for computing an MMDC matching between $A$ and $B$ with total cardinality $n$. In fact, we modified the basic Hungarian algorithm to get a new algorithm, called the MMDC matching algorithm. Then, we construct a bipartite graph $G$ and apply our new algorithm on $G$.

\end{document}